\documentclass[twocolumn,pre,floatfix,showkeys,showpacs]{revtex4}
\usepackage[colorlinks,linkcolor={blue},citecolor={blue},urlcolor={blue}]{hyperref}

\usepackage[]{graphicx}

\begin{document}

\title{Equilibrium and non-equilibrium fluctuations in a glass-forming liquid}

\author{ Azita Parsaeian and Horacio E. Castillo }
\affiliation{ Department of Physics and Astronomy, Ohio University,
  Athens, OH, 45701, USA } 
\date{November 17, 2008} 

\begin{abstract}
Glass-forming liquids display strong fluctuations -- dynamical
heterogeneities -- near their glass transition. By numerically
simulating a binary Weeks-Chandler-Andersen liquid and varying both
temperature and timescale, we investigate the probability
distributions of two kinds of local fluctuations in the
non-equilibrium (aging) regime and in the equilibrium regime; and
find them to be very similar in the two regimes and across
temperatures. We also observe that, when appropriately rescaled, the
integrated dynamic susceptibility is very weakly dependent on
temperature and very similar in both regimes.
\end{abstract}

%
%
%


\maketitle

The study of fluctuations in statistical mechanics has had a long
history, starting from Einstein's seminal work on Brownian motion.
Recent research on fluctuations has focused on nonequilibrium
problems, such as
turbulence~\cite{bramwell-holdsworth-pinton_universal-rare-fluctuations-turbulence-critical_nature-396-552-1998},
non-equilibrium steady state
phenomena~\cite{taniguchi-cohen_noneq-steady-state-therm-fluct-stoch_jstatphys-130-633-2008},
single molecule experiments in
biophysics~\cite{ritort_noneq-therm-small_crphys-8-528-2007}, or the
relaxation of large perturbations~\cite{fdt-large-amplitude}.  In the
present work, we want to use glasses as a convenient laboratory to
study fluctuations in both equilibrium and out of equilibrium states;
{\em i.e.\/} on both sides of the (kinetic) glass ``transition''.

As experimentally observed, the glass transition is not a sharp
thermodynamic phase transition~\cite{thermodynamic_phase_transition},
but rather a dynamical crossover. As the system approaches this
crossover, the equilibration time $\tau_{\mbox{\scriptsize eq}}$
gradually increases, until it eventually becomes longer than the
representative laboratory timescale $t_{\mbox{\scriptsize
    lab}}$~\cite{glass_transition,
  ediger_hetdyn_review_arpc-51-99-2000}. Beyond the glass transition,
the system cannot equilibrate in laboratory timescales. This leads to
the emergence of new phenomena, such as physical aging, {\em i.e.\/}
the breakdown of time translation invariance (TTI). Besides the
slowdown of the dynamics, also dynamical
heterogeneities~\cite{ediger_hetdyn_review_arpc-51-99-2000,
  sillescu_hetdyn-review_jncs-243-81-1999}, {\em i.e.\/} strong local
fluctuations in the properties of glass-forming liquids, are
observed
both in the equilibrium
regime~\cite{Kegel-Blaaderen-science00, Weeks-Weitz,
  Glotzer_simulations} and in the aging
regime~\cite{Courtland-Weeks-jphysc03,
  sinnathamby-oukris-israeloff_prl05, parisi_99,
  castillo-parsaeian_shortrhoC_naturephys-3-26-2007,
  parsaeian-castillo_shortcorr_condmat-0610789,
  parsaeian-castillo_polymer_arxiv_submitted}. It is not clear,
though, to what degree the properties of those fluctuations depend on
whether they are observed in one or the other 
regime.  In the present work, we address this question in the context
of a detailed molecular dynamics simulation of a simple glass model
probing the aging regime, the equilibrium regime, and the crossover
regime between the two.

A theoretical framework based on the presence of a Goldstone mode in
the aging dynamics~\cite{rpg} that gives rise to local fluctuations in
the age of the sample~\cite{ckcc,
  chamon-charb-cug-reich-sellito_condmat04} predicts that probability
distributions of local fluctuations in the aging regime are
approximately independent of the waiting time $t_w$ at fixed values of
the two-time global correlation $C_{\mbox{\scriptsize
    global}}(t,t_w)$. This prediction, initially proposed for spin
glasses~\cite{ckcc}, has been found to apply also to off-lattice models
of aging structural
glasses~\cite{castillo-parsaeian_shortrhoC_naturephys-3-26-2007,
  parsaeian-castillo_polymer_arxiv_submitted} (where
$C_{\mbox{\scriptsize global}}$ is the self part of the intermediate
scattering function: $C_{\mbox{\scriptsize global}}(t,t_w) \equiv
\frac{1}{N}\sum_{j=1}^{N}\exp(i{\bf q}.({\bf r_j}(t)-{\bf r_j}(t_w)))$
). Additionally, it has been found that the dynamic spatial
correlations in an aging structural glass display a simple scaling
behavior as a function of $C_{\mbox{\scriptsize
    global}}$~\cite{parsaeian-castillo_shortcorr_condmat-0610789,
  parsaeian-castillo_polymer_arxiv_submitted}. In what follows we
address the question of whether these same scaling behaviors extend to
the equilibrium regime of an off-lattice structural glass model.

\begin{figure}[ht]

  \begin{center}
    \includegraphics[width=3.5in]{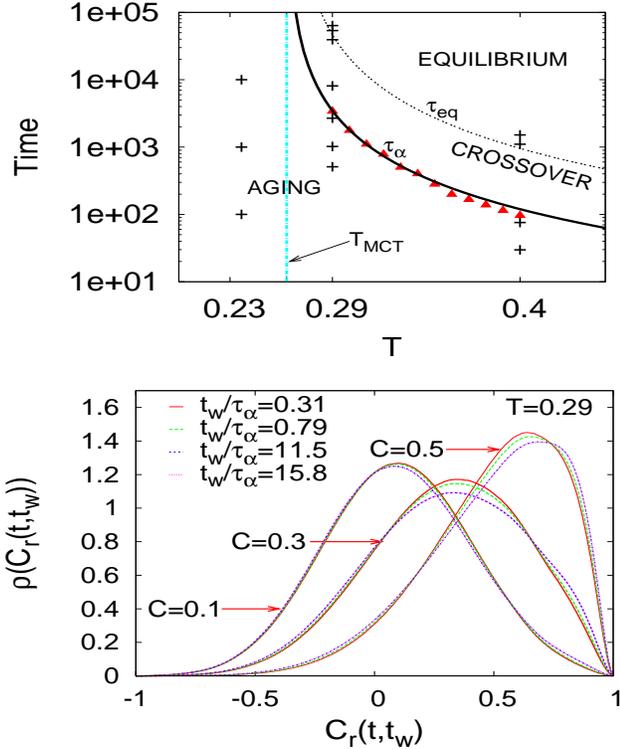}

    \caption{{\em Top panel:} Timescales as functions of the
      temperature.  ``+'' symbols: (temperature, waiting time) pairs
      for which we show our results. Triangles: $\alpha$ relaxation
      times for $T=0.29,\cdots,0.40$. Full line: fit of
      $\tau_{\alpha}$ to $\tau_{\alpha} =
      a(T-T_{\mbox{\scriptsize MCT}})^{-\gamma}$. Dotted line: $\tau_{\mbox{\scriptsize
          eq}}$, estimated as $\tau_{\mbox{\scriptsize eq}}\sim 10
      \tau_{\alpha}$ (see text). Vertical dash-dotted line:
      $T_{\mbox{\scriptsize MCT}}$. {\em Bottom panel:} Probability distribution
      $\rho(C_{\bf r})$ at $T=0.29$, when $C_{\mbox{\scriptsize
          global}}(t,t_w)=0.1,0.3, 0.5$, for $t_w/\tau_{\alpha} =
      0.31, 0.79$ (aging) and $t_w/\tau_{\alpha} = 11.5, 15.8$
      (equilibrium). The two equilibrium plots overlap perfectly. The
      coarse graining region contains on average $6.6$ particles.}
    \label{fig:regions-prob-cdr}

  \end{center}

\end{figure}

\begin{figure}[ht]
\begin{center} 
\includegraphics[width=3.5in]{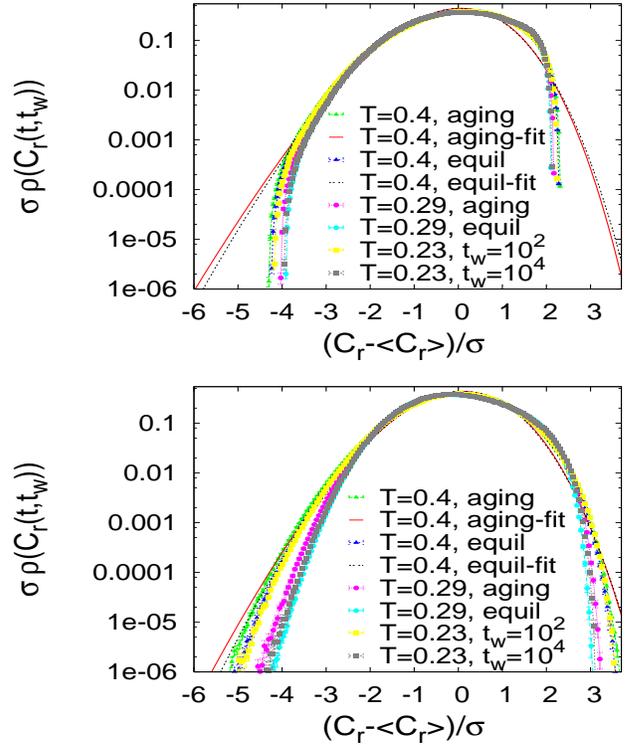}

    \caption{Rescaled probability distributions $\sigma_C \rho(C_{\bf
        r})$ as functions of the normalized fluctuation $(C_{\bf
        r}-C_{\mbox{\scriptsize global}})/\sigma_C$ in the two-time
      correlation, shown at $t_w/\tau_{\alpha} = 0.79$ (aging regime)
      and $t_w/\tau_{\alpha} = 11.5$ (equilibrium regime), for
      temperatures $T=0.4$ (triangles), $T=0.29$ (circles) and
      $T=0.236$ (squares), with $C_{\mbox{\scriptsize global}}(t,t_w)
      = 0.3$. Generalized Gumbel fits to the distributions for $T=0.4$
      are shown with thin lines. {\em Top panel:} Coarse graining
      region containing on average $6.6$ particles/box. {\em Bottom
        panel:} Coarse graining region containing on average $30.5$
      particles/box. }
    \label{fig:prob-cdr-gumb}

\end{center} 
\end{figure}

\begin{figure}[ht]
  \begin{center} 
    \includegraphics[width=3.5 in]{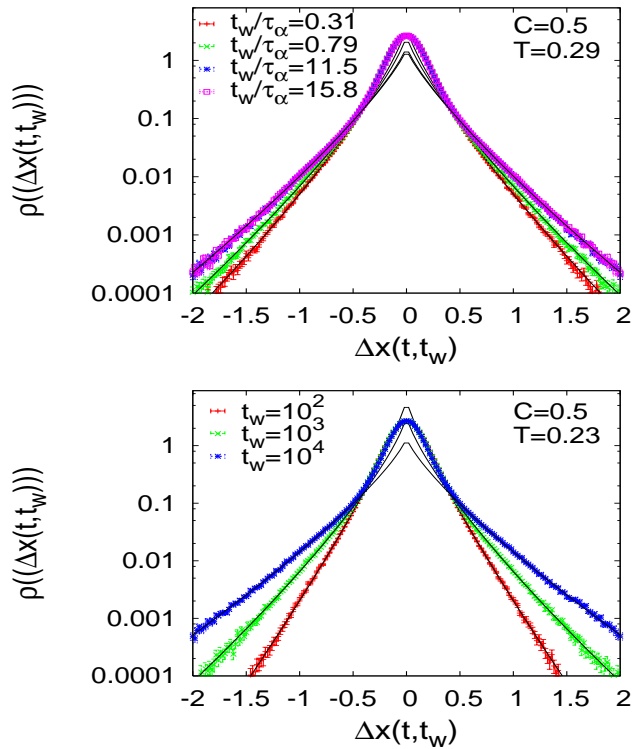}
    \caption{ $\rho(\Delta x)$ for $C_{\mbox{\scriptsize
          global}}(t,t_w)=0.5$, plotted with a logarithmic vertical
      axis to emphasize the tails of the distributions. Fits to the
      tails by a nonlinear exponential form are shown with full lines.
      {\em Top panel:} $T=0.29$. $t_w = 0.31 \, \tau_{\alpha}, 0.79
      \, \tau_{\alpha}$ are in the aging regime; whilst $t_w =
      11.5 \, \tau_{\alpha}, 15.8 \, \tau_{\alpha}$, which give perfectly
      overlapping results, are in the equilibrium regime. {\em Bottom
        Panel:} $T=0.23$. Only the aging regime is accessible; $t_w =
      10^2, 10^3, 10^4$. }
    \label{fig:prob-dxi}
  \end{center} 
\end{figure}

\begin{figure}[ht]
\begin{center} 
\includegraphics[width=3.5 in]{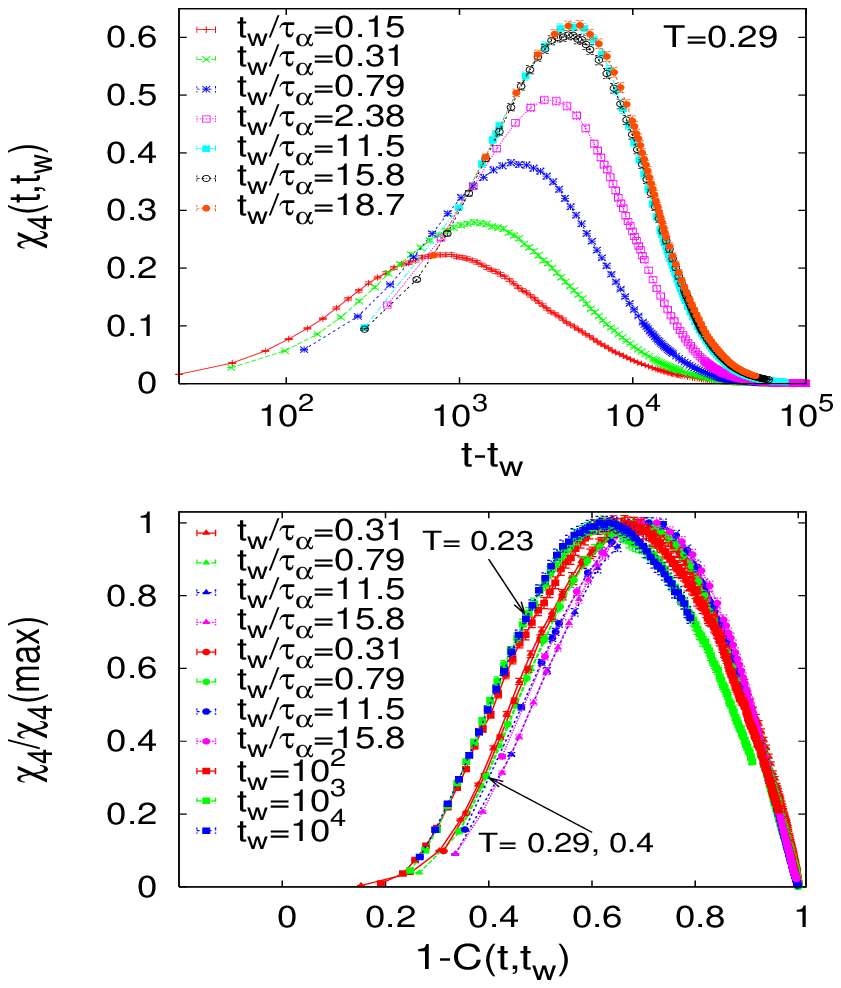}
    \caption{{\em Top panel:} $\chi_4(t,t_w)$ as a function of $t-t_w$
      for $T=0.29$. {\em Bottom panel:} $\chi_4$ divided by its
      maximum value $\chi_4(\mbox{max})$, plotted as a function of
      $1-C$, with $C \equiv C_{\mbox{\scriptsize global}}(t,t_w)$ for
      $T=0.4$ (triangles), $T=0.29$ (circles) and $T=0.23$ (squares).}
    \label{fig:chi4}
  \end{center}
\end{figure}

In order to observe both equilibrium and aging phenomena, one could
either reduce the temperature $T$ of the thermal reservoir that the
system is coupled to beyond the glass transition temperature $T_g$ at
a fixed laboratory timescale $t_{\mbox{\scriptsize lab}}$, or reduce
$t_{\mbox{\scriptsize lab}}$ until it becomes shorter than
$\tau_{\mbox{\scriptsize eq}}$, at fixed reservoir temperature $T$.
Here we use both approaches. To observe genuinely glassy dynamics, we
make sure that $T$ is low enough that two-step relaxation is observed,
and that timescales are much longer than the timescale of the first
step. 

We simulate a $80:20$ binary mixture of particles of mass $M_A = M_B =
M$, in three dimensions, at a number density $\rho = 1.204$,
interacting via short-range, purely repulsive Weeks-Chandler-Andersen
(WCA) potentials~\cite{weeks-chandler-andersen}: $ V_{\alpha,\beta} =
4 \, \epsilon_{\alpha,\beta} \, \left[
  \left(\frac{\sigma_{\alpha,\beta}}{r}\right)^{12}
  -\left(\frac{\sigma_{\alpha,\beta}}{r}\right)^6 +1 \right]
\theta(r^{(\mbox{\scriptsize cutoff})}_{\alpha,\beta} -r)$, where
$\alpha,\beta \in \{A,B\}$ denotes the particle type,
$\epsilon_{\alpha,\beta}$ and $\sigma_{\alpha,\beta}$ are taken as
in~\cite{Kob-Barrat-aging-prl97, Kob-Andersen}, $\theta(x)$ is the
Heaviside theta function, and $r^{(\mbox{\scriptsize
    cutoff})}_{\alpha,\beta} \equiv 2^{1/6} \sigma_{\alpha,\beta}
\approx 1.1225\;\sigma_{\alpha, \beta}$ makes $V(r)$ and $\partial V /
\partial r$ continuous for all $r>0$.  We take $k_{B}=1.0$, and the
units of length, energy and time as $\sigma_{AA}$, $\epsilon_{AA}$ and
$\tau_{LJ} \equiv {({\sigma_{AA}}^2 M/48 \epsilon_{AA})}^{1/2}$
respectively, where
$\tau_{LJ}$ represents roughly the timescale for vibrational motion in
the system. Initially the system is equilibrated at a high
temperature: $T_i=5.0$, and then instantaneously quenched to a final
temperature $T_f  \ll T_i$.
The top panel of Fig.~\ref{fig:regions-prob-cdr} illustrates the
temperature and time regimes for which we show results. We take
three different $T_f$'s. For $T_f=0.236 \approx 0.9 \,
T_{\mbox{\scriptsize MCT}}$, only the aging regime is observed within
our simulation time frame. For $T_f=0.29 \approx 1.1 \,
T_{\mbox{\scriptsize MCT}}$ and $T_f =0.4 \approx 1.5 \,
T_{\mbox{\scriptsize MCT}}$, we observe both the aging and equilibrium
regimes. Here, $T_{\mbox{\scriptsize MCT}}$ denotes the Mode Coupling
Theory (MCT) critical temperature. We define $\tau_{\alpha}$ by the
condition $\lim_{t_w \to \infty} C(t_w+\tau_{\alpha},t_w) =1/e $ and
find it for the temperatures $T=0.29,0.30 \cdots 0.4$.
We observe that for $2.9 \le T \le 3.4$, the equilibrium $C(t,t_w)$
displays two step relaxation and also scaling behavior as predicted by
MCT. Following~\cite{Kob-Andersen}, we fit
$\tau_{\alpha}=(T-T_{\mbox{\scriptsize MCT}})^{-\gamma}$ in that
range, and find $T_{\mbox{\scriptsize MCT}}=0.263 \pm 0.01$ and
$\gamma=2.1 \pm 0.7$. 

The time of the quench is taken as the origin of times $t=0$. In our
simulations, $t_{\mbox{\scriptsize lab}}$ is replaced by $t_w$. We
find that aging effects are strongest for $t_w/\tau_{\alpha} \lesssim
2$. For $2 \lesssim t_w/\tau_{\alpha} \lesssim 10$ the aging effects
gradually become weaker with increasing $t_w$ until they disappear for
$t_w/\tau_{\alpha} \sim 10$. We thus estimate $\tau_{\mbox{\scriptsize
    eq}} \sim 10 \tau_{\alpha} $. We show results for a $1,000$
particle system, with $5000$, $9000$ and $10,000$ independent thermal
histories when $T_f=0.236$, $T_f=0.29$, and $T_f=0.4$
respectively. 

We present results for the probability distributions of
observables which probe local fluctuations in small
regions of the system: the local coarse grained two-time correlation
function~\cite{castillo-parsaeian_shortrhoC_naturephys-3-26-2007}
$C_{{\bf r}}(t,t_w) \equiv \frac{1}{N(B_{\bf r})} \sum_{{\bf r}_j(t_w) \in B_{\bf r}} 
  \cos({\bf q} \cdot ({\bf r}_j(t) - {\bf r}_j(t_w)))$,
and the particle displacements $\Delta x_j(t,t_w) = x_j(t) - x_j(t_w)$
along one direction~\cite{Weeks-Weitz,
  castillo-parsaeian_shortrhoC_naturephys-3-26-2007,
  chaudhuri-gao-berthier-kilfoil-kob_random-walk-hetdyn-attracting-colloid_arxiv-07120887}. 
Here $B_{\bf r}$ represents a small coarse graining box around the
point ${\bf r}$ in the system and $N(B_{\bf r})$ is the number of
particles {\em present at the waiting time $t_w$} in the box $B_{\bf
r}$. We choose a value of q that corresponds to the main peak in the
structure factor $S(q)$ of the system, $q = 7.2$.
In order to probe the spatial correlations of the fluctuations, we
also consider the generalized $4$-point density susceptibility $\chi_4
\equiv \int {d^3{\bf r}} \; g_4({\bf r}, t,
t_w)$~\cite{Glotzer_simulations, Biroli_etal,  
parsaeian-castillo_shortcorr_condmat-0610789}, where $g_4({\bf r}, t,
t_w)$ is a $4$-point ($2$-time,
$2$-position) correlation function~\cite{Glotzer_simulations, 
parsaeian-castillo_shortcorr_condmat-0610789}.

In the bottom panel of Fig.~\ref{fig:regions-prob-cdr} we show the
probability distributions $\rho(C_{\bf r})$ for $C_{\mbox{\scriptsize
    global}}(t,t_w) = 0.1, 0.3, 0.5$.
We observe that, for fixed value of $C_{\mbox{\scriptsize
    global}}(t,t_w)$, the probability distributions are approximately
invariant between the two regimes.  These results extend those found
for the aging regime of a binary Lennard-Jones (LJ)
glass~\cite{castillo-parsaeian_shortrhoC_naturephys-3-26-2007}.

Fig.~\ref{fig:prob-cdr-gumb} shows the rescaled probability
distributions $\sigma_C \rho(C_{\bf r})$ versus the normalized
fluctuation $(C_{\bf r}-C_{\mbox{\scriptsize global}})/\sigma_C$ in
the one-point two-time correlator, for coarse graining regions of two
sizes, 
for the temperatures
$T=0.23$, $T=0.29$ and $T=0.4$. We observe a very good collapse of the
data for different $t_w$ and different temperatures for the smaller
coarse graining size, and a slightly less good collapse for the
larger coarse graining size. This suggests that the weak dependence
on $t_w$ may be due to the time dependence of the dynamic correlation
length~\cite{castillo-parsaeian_shortrhoC_naturephys-3-26-2007}.
In~\cite{chamon-charb-cug-reich-sellito_condmat04} it has been argued
that the probability distribution of the normalized fluctuations
$(C_{\bf r}-C_{\mbox{\scriptsize global}})/\sigma_C$ can be described
well by a generalized Gumbel distribution $\Phi_{\mbox {\scriptsize
    Gumbel}}$~\cite{bramwell-holdsworth-pinton_universal-rare-fluctuations-turbulence-critical_nature-396-552-1998}. 
Fig.~\ref{fig:prob-cdr-gumb} shows that fits of our data to
$\Phi_{\mbox{\scriptsize Gumbel}}$ become better for a larger coarse
graining region, i.e. when averaging affects have started to modify
the
distribution~\cite{bramwell-holdsworth-pinton_universal-rare-fluctuations-turbulence-critical_nature-396-552-1998}.

In Fig.~\ref{fig:prob-dxi} we show the probability distribution
$\rho(\Delta x)$ of the particle displacements, with
$C_{\mbox{\scriptsize global}}(t,t_w) = 0.5$, for both $T_f = 0.29 >
T_{\mbox{\scriptsize MCT}}$ (top panel) and $T_f = 0.23 <
T_{\mbox{\scriptsize MCT}}$ (bottom panel). We use semilog plots to
emphasize the tails of the distributions. The distributions are very
similar to those determined by confocal microscopy for an equilibrium
sterically stabilized colloidal liquid near its glass
transition~\cite{Weeks-Weitz}, and also to the ones found in
simulations of an aging binary LJ
glass~\cite{castillo-parsaeian_shortrhoC_naturephys-3-26-2007}; but
markedly different from those found by confocal microscopy in an
attracting
colloid~\cite{chaudhuri-gao-berthier-kilfoil-kob_random-walk-hetdyn-attracting-colloid_arxiv-07120887}.
For small to moderate values of $\Delta x$, the data for different
$t_w$ and temperatures collapse very well with each other and with a
common gaussian fit, as long as $C_{\mbox{\scriptsize global}}$ is
constant (not shown).
We find, however, that for fixed $C_{\mbox{\scriptsize
    global}}(t,t_w)$, the tails of the distribution become wider for
increasing $t_w$. The tails can be fit in the region $|\Delta x | >
0.5$ by a nonlinear exponential form $\rho(\Delta x ) \approx N \exp(-
{|\Delta x/a|}^{\beta} )$, with $\beta \sim 1$ decreasing
monotonically with increased
$t_w$~\cite{castillo-parsaeian_shortrhoC_naturephys-3-26-2007}. For the case of $T >
T_{\mbox{\scriptsize MCT}}$ both the tails and the value of $\beta$
gradually saturate as the system approaches equilibrium.

To quantify the spatial correlation of dynamical fluctuations, the
4-point density susceptibility
$\chi_4(t,t_w)$~\cite{Glotzer_simulations, Biroli_etal,
  parsaeian-castillo_shortcorr_condmat-0610789} and similar quantities
have been used both in numerical
simulations~\cite{Glotzer_simulations, Biroli_etal,
  parsaeian-castillo_shortcorr_condmat-0610789} and in
experiments~\cite{mayer-bissig-berthier-cipelletti-garrahan-sollich-trappe_heterogenous-dynamics-coarsening_prl-93-115701-2004,
  experiments-supercooled_chi4, experiments-granular_chi4,
  duri-cipelletti_lengthscale-dynhet-colloidal-gel_epl-76-972-2006}. The
top panel of Fig.~\ref{fig:chi4} shows $\chi_4(t,t_w)$ as a function
of $t-t_w$ at  $T_f = 0.29$, for various $t_w$.
We have found
that $\chi_4(t,t_w)$ has a peak whose height and location grow with
$t_w$ as long as the system is aging. This behavior as $t_w$ is increased is observed experimentally in a coarsening
foam~\cite{mayer-bissig-berthier-cipelletti-garrahan-sollich-trappe_heterogenous-dynamics-coarsening_prl-93-115701-2004},
and analogous to the behavior of supercooled liquids as the
temperature is reduced~\cite{experiments-supercooled_chi4} and of
granular systems as the area fraction is
increased~\cite{experiments-granular_chi4}.
In the bottom panel of Fig.~\ref{fig:chi4} we test for a
possible scaling behavior with $C \equiv C_{\mbox{\scriptsize
    global}}$, by plotting the rescaled quantity
$\chi_4/\chi_4(\mbox{max})$ as a function of $1-C$, 
for $T_f =0.23, 0.29, 0.4$. An exact collapse would implicate the
factorization $\chi_4(t,t_w) = \chi_4^\circ(t_w)\phi(C(t,t_w))$, where
$\chi_4^\circ(t_w) \equiv \chi_4(t,t_w)|_{C(t,t_w)=1/e}$ is a
rescaling factor that depends only on $t_w$ until it saturates when
the system equilibriates; and $\phi(C(t,t_w))$ is a scaling function
which depends on times only through the value of the intermediate
scattering function $C(t, t_w)$. We observe that the curves
approximately collapse into two groups, one for $T >
T_{\mbox{\scriptsize MCT}}$ and another for $T < T_{\mbox{\scriptsize
    MCT}}$, with a slight horizontal shift between the two
groups. Curves for the same temperature approximately collapse,
although a slight systematic variation with $t_w$ remains.

In summary, we have analyzed the probability distributions of local
two-time observables and the spatial correlations of fluctuations in
the equilibrium and aging regimes in a simple glass model.
We do find some differences between the two regimes. For example, the
tails of the probability distributions of one-particle displacements
$\rho(\Delta x)$ become slightly wider as $t_w$ grows at constant
$C_{\mbox{\scriptsize global}}(t,t_w)$. Also, the rescaled generalized
susceptibility $\chi_4/\chi_4(\mbox{max})$, when plotted as a function
of $1- C_{\mbox{\scriptsize global}}$, shows slight systematic
dependences on $t_w$ and on $T$.
However, those differences are minor, and the similarities are quite
striking. In general terms, the simple scaling behaviors previously
observed in the aging
regime~\cite{castillo-parsaeian_shortrhoC_naturephys-3-26-2007,
  parsaeian-castillo_shortcorr_condmat-0610789} extend to the
equilibrium regime. On one hand, both the probability distributions of
local two-time correlations and the probability distributions of
one-particle displacements are approximately invariant between the two
regimes as long as the total correlation $C_{\mbox{\scriptsize
    global}}(t,t_w)$ is kept constant. On the other hand, the rescaled
dynamic susceptibility $\chi_4/\chi_4(\mbox{max})$, when plotted
against $1-C_{\mbox{\scriptsize global}}(t,t_w)$, is also
approximately invariant between the two regimes. Additionally, we find
that the probability distributions $\rho(C_{\bf r})$ of local two-time
correlations can be fitted by generalized Gumbel
forms~\cite{chamon-charb-cug-reich-sellito_condmat04,
  bramwell-holdsworth-pinton_universal-rare-fluctuations-turbulence-critical_nature-396-552-1998},
as long as the coarse graining size is relatively large. Finally, the
scaling functions describing the distribution of local two-time
correlations and the dynamic susceptibilities are very weakly
dependent on temperature across the range from $T \lesssim
T_{\mbox{\scriptsize MCT}}$ to $T > T_{\mbox{\scriptsize MCT}}$, and
qualitatively similar to those found in an aging LJ
glass~\cite{castillo-parsaeian_shortrhoC_naturephys-3-26-2007,
  parsaeian-castillo_shortcorr_condmat-0610789}.

H.~E.~C. thanks L.~Berthier, J.~P.~Bouchaud, L.~Cugliandolo,
S.~Glotzer, N.~Israeloff, M.~Kennett, M.~Kilfoil, D.~Reichman, 
E.~Weeks, and particularly C.~Chamon for suggestions and discussions.
This work was supported in part by DOE under grant DE-FG02-06ER46300,
by NSF under grant PHY99-07949, and by Ohio University. Numerical
simulations were carried out at the Ohio Supercomputing
Center. H.~E.~C. acknowledges the hospitality of the Aspen Center for
Physics, were part of this work was performed.


\begin{thebibliography}{99}
\bibitem{thermodynamic_phase_transition}
  However, such transition may exist at a lower
  temperature~\cite{glass_transition}.
\end{thebibliography}
\end{document}